\documentclass{article}

\usepackage[dblblindworkshop, final]{neurips_2026}

\workshoptitle{TAE (Trust-AI-Eval): Can We Trust AI Evaluation?}

\usepackage{tcolorbox}      
\usepackage{wrapfig}
\usepackage[utf8]{inputenc} 
\usepackage[T1]{fontenc}    
\usepackage{hyperref}       
\usepackage{url}            
\usepackage{booktabs}       
\usepackage{amsfonts}       
\usepackage{nicefrac}       
\usepackage{microtype}      
\usepackage{xcolor}         
\usepackage{tikz}
\usetikzlibrary{positioning,arrows.meta}
\usepackage[utf8]{inputenc} 
\usepackage{amssymb}

\usepackage{microtype}      
\usepackage{amsmath,amssymb}
\usepackage{graphicx}
\usepackage{booktabs}
\usepackage{enumitem}
\usepackage{hyperref}
\usepackage{url}
\usepackage{enumitem}
\usepackage{placeins}       
\usepackage{tcolorbox}
\tcbuselibrary{skins}
\definecolor{exampleblue}{RGB}{240, 245, 255}

\newtcolorbox{examplebox}[1]{
  colback=exampleblue,
  colframe=black!70,
fontupper=\scriptsize,
fonttitle=\scriptsize\bfseries\sffamily,  title=Example: #1,
  left=6pt, right=6pt, top=4pt, bottom=4pt,
  boxrule=0.8pt,
  arc=3pt
}
\usepackage{array}
\newcolumntype{L}[1]{>{\raggedright\arraybackslash\hspace{0pt}}p{#1}}

\title{Evaluating Deception, Not Just Deepfakes}

\author{%
  Jessee Ho \\
  Vector Institute\\
  Toronto, Ontario, Canada \\
 \texttt{jessee.ho}\\
 \texttt{@queensu.ca} \\
   \And
  Shweta Khushu \\
  Vector Institute\\
  Toronto, Ontario, Canada \\
  \texttt{shweta.khushu}\\
  \texttt{@vectorinstitute.ai} \\
  \And
  Shaina Raza \\
  Vector Institute\\
  Toronto, Ontario, Canada \\
 \texttt{shaina.raza}\\
 \texttt{@vectorinstitute.ai} \\
}

\begin{document}

\maketitle

\begin{abstract}
Deepfake detection is typically evaluated as a media-classification task: given a clip, determine whether it is synthetic. Yet benchmark performance is increasingly invoked to support broader claims about preventing deepfake-enabled deception in video calls, voice calls, and targeted communications. We argue that these claims require evidence across two distinct validity gaps. \textbf{External validity} concerns whether detection performance transfers across generators, processing pipelines, and communication channels. \textbf{Consequential validity}, as used here, concerns whether reliable detection translates into timely intervention and prevention of harmful actions. Even perfect synthetic-origin classification does not guarantee prevention when verification is unavailable, intervention occurs too late, or recipients disregard warnings. We therefore propose a scenario-level evaluation framework that integrates media evidence, interaction-level indicators grounded in speech act theory, conversational pragmatics, and influence principles, and operational outcomes. We define candidate measures of attack prevention, benign pass-through, intervention precision under explicit deployment base rates, and intervention timing. Our position is that deployment-oriented evaluation must measure not only whether synthetic media is detected, but whether deception is prevented without unnecessarily disrupting legitimate interactions \href{https://jesseeho.github.io/deepfake-deception/}{(project page)}.
\end{abstract}

\section{Introduction}
\label{sec:introduction}

\begin{wrapfigure}{r}{0.55\textwidth}
\vspace{-10pt}
\centering
\includegraphics[width=\linewidth]{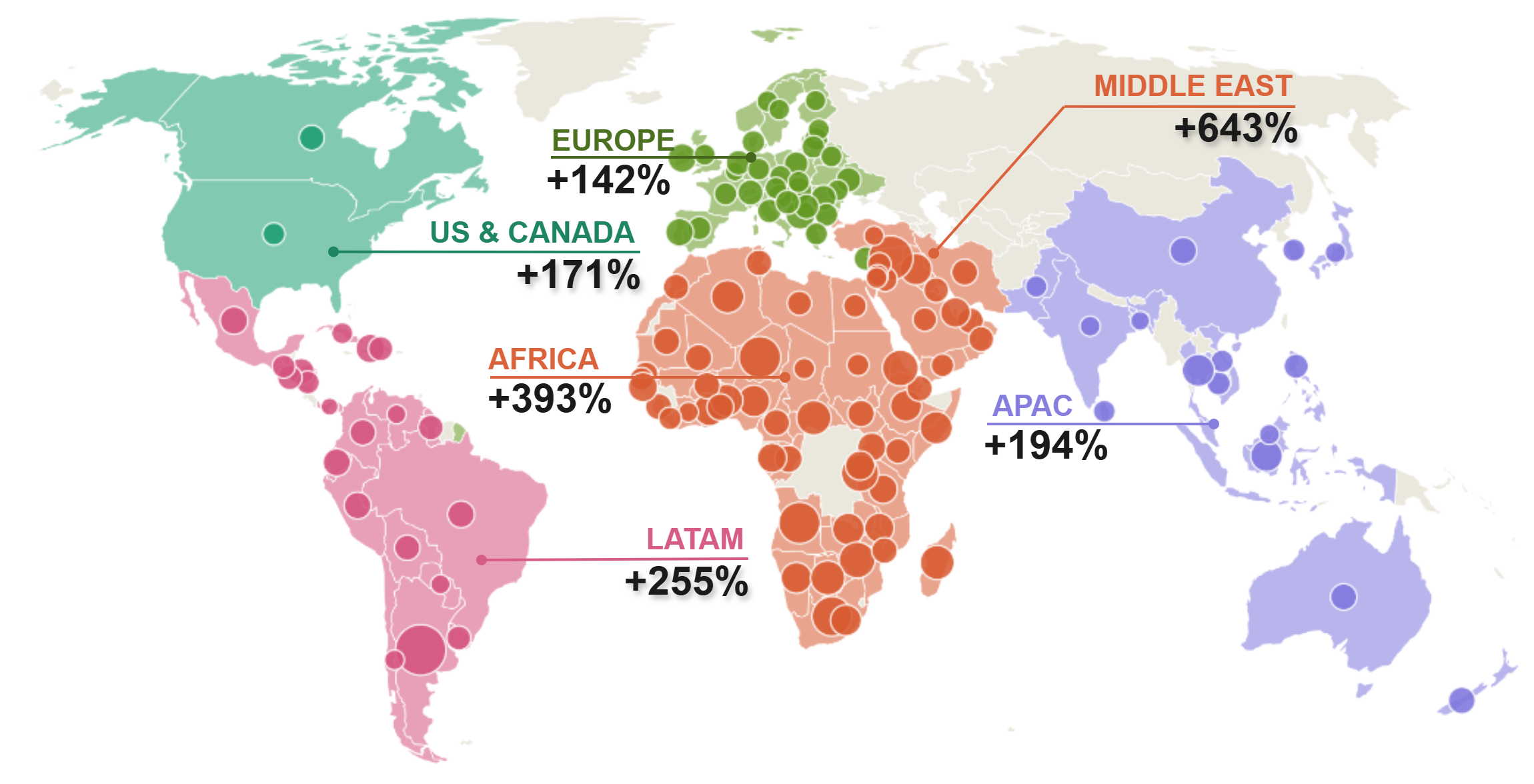}
\caption{Global growth in deepfake-driven identity fraud from 2023 to 2024 (in \%). Circle size scales with country-level growth in identity fraud, while regional labels highlight growth in deepfake-specific fraud~\cite{sumsub2024identityfraud}.}
\label{fig:deepfake_map}
\vspace{-15pt}
\end{wrapfigure}

Deepfake \footnote{The term `deepfake'' is widely believed to have originated on Reddit in 2017, where a user (`deepfakes'') shared AI-generated face-swapped videos~\cite{bbc_reddit}.} detection presents a growing mismatch between benchmark success and real-world reliability. State-of-the-art detectors report strong performance on curated datasets such as FaceForensics++~\cite{rossler2019faceforensics}, Celeb-DF~\cite{li2020celeb}, and DeeperForensics~\cite{jiang2020deeperforensics}, yet recent cross-distribution evaluations show substantial performance degradation on contemporary generators and real-world media~\cite{chandra2025deepfakeeval,Le2024SoKSA}. At the same time, the operational stakes are rising: deepfake-enabled fraud incidents increased roughly 40-fold between 2022 and 2024~\cite{sumsub2023identityfraud,sumsub2024identityfraud}, while generative-AI-facilitated fraud losses in the United States alone are projected to reach \$40 billion annually by 2027~\cite{deloitte2024genai}. Figure~\ref{fig:deepfake_map} illustrates the corresponding global growth of deepfake-enabled fraud~\cite{sumsub2024identityfraud}. Together, these trends expose a central challenge: strong benchmark accuracy does not necessarily translate into reliable real-world detection.

Benchmarks usually measure whether a model can identify synthetic media, while real-world deployment requires more. This reflects a broader issue in ML evaluation: benchmark results support wider claims only when the benchmark measures what those claims require~\cite{raji2021benchmark,bowman2021benchmarkNLP,freiesleben2025benchmarking}, raising a basic question of evaluation validity: \emph{does the benchmark measure the properties needed to support claims about real-world deepfake detection?} 

This question matters most in high-stakes, real-time interactions described above, where misjudgment can cause harm. We therefore propose to focus on \emph{interactive deepfakes}: AI generated audio-visual impersonation used as a \textbf{social-engineering mechanism} in video calls, voice calls, or targeted communications~\cite{pei2026deepfake}, where a recipient must decide whether to trust, comply with, or act on a communication. This setting is increasingly important because voice and video have traditionally served as strong identity cues in remote communication. As both can now be convincingly synthesized, users may need to judge authenticity from how the interaction unfolds, not only from how the caller looks or sounds.

Reported incidents already illustrate this shift from synthetic-media generation to interactive social engineering.  In a 2026 Japanese case, a suspected deepfake job applicant withdrew  when in-person attendance was required, and later checks linked the  submitted identity to a real executive~\cite{unitedpress2026, USGov2026NKdeepfake}.  In an attempted impersonation of a Ferrari executive, the recipient  stopped the attack using a shared-knowledge identity  check~\cite{bloomberg2024ferrari}. In both cases, protection depended on information beyond the audio-visual signal itself.

We do not claim that deception prevention is the appropriate objective for all synthetic-media applications. Standalone media classification remains important for content moderation, provenance, and forensic analysis. Rather, our claim is that, in interactive impersonation and fraud, identifying synthetic origin and preventing deception are distinct evaluation constructs.
Building on prior work on benchmark and measurement validity \cite{freiesleben2025benchmarking, raji2021benchmark}, we distinguish two \textbf{validity gaps}: \textbf{external validity}, whether benchmark performance transfers across generators, platforms, compression, and communication channels; and \textbf{consequential validity}, whether successful deepfake detection actually contributes to preventing deception. Existing benchmarks primarily measure detection performance under benchmark conditions, providing limited evidence for either form of validity. These gaps motivate evaluating not only the media itself, but also the interaction and decision context in which deception occurs.

\textbf{Contributions:}  We develop this position in three steps:

\begin{enumerate}[leftmargin=*, itemsep=1pt, topsep=2pt, parsep=0pt, partopsep=0pt]
    \item We distinguish two missing validity gaps (external and consequential) between benchmark accuracy and real-world deception prevention and our analysis shows that even perfect synthetic-media classification would leave the consequential gap unresolved (\S\ref{sec:validity-gaps}).
    \item We operationalize \textbf{interactive deception} as an evaluation construct and derive three candidate interaction-level indicators: act validity, conversational coherence, and influence patterns, grounded in Speech Act Theory, Gricean pragmatics, and Cialdini's principles of influence (\S\ref{sec:interaction-evidence}).
    \item We translate this reframing into a \textbf{scenario-level evaluation protocol} that evaluates the full interaction rather than an isolated media clip, using attack prevention rate, benign pass-through rate, precision under stated deployment base rates, and intervention latency (\S\ref{sec:evaluation}).
\end{enumerate}

\begin{figure*}[h]
\vspace{-9pt}
\centering
\includegraphics[width=0.78\textwidth]{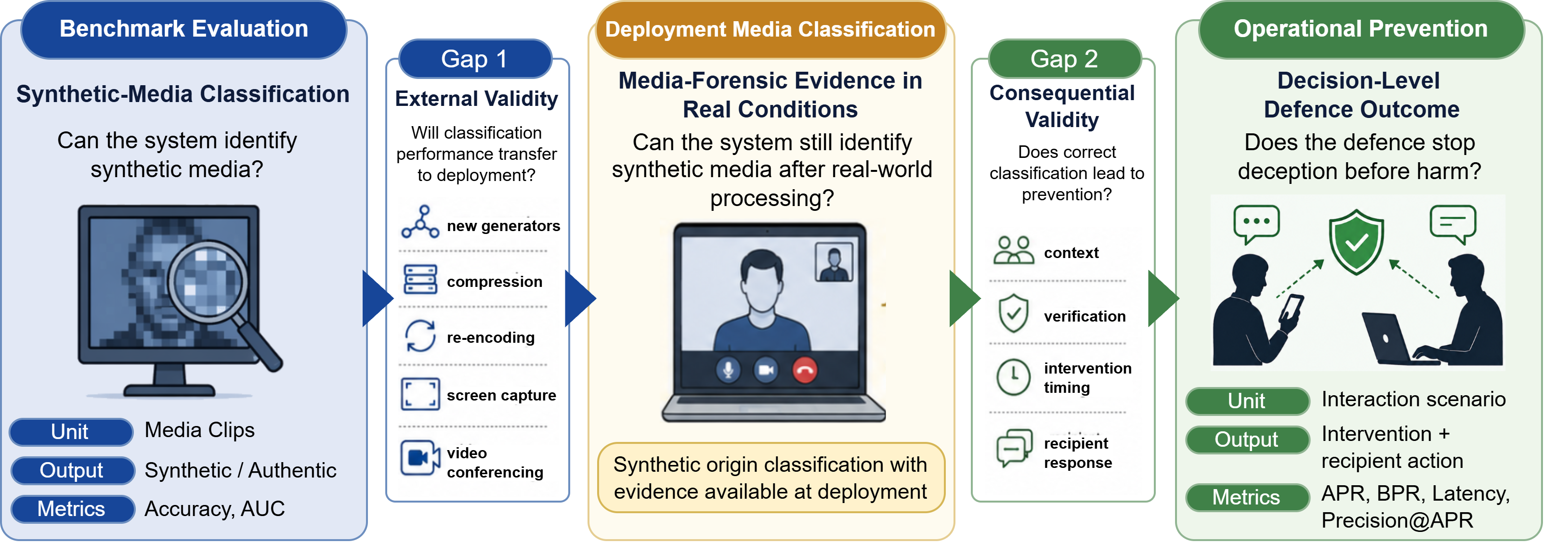}
\caption{From classification to prevention. Deepfake benchmarks test whether media is synthetic. Deployment claims require two further checks: external validity, whether performance transfers to real-world conditions, and construct/consequential validity, whether correct classification supports effective prevention. The framework therefore combines media forensics with interaction-level, decision-aware evaluation.}
\label{fig:interrogation}
\end{figure*}
\vspace{-5pt}

\section{Validity Gaps: From Benchmark Performance to Prevention}
\label{sec:validity-gaps}
Moving from benchmark performance to a claim of real-world deception prevention requires two distinct validity steps. 
We examine each gap in turn.

\subsection{External Validity: From Benchmark to Deployment}
\label{sec:external-validity}
The first step from deepfake benchmark performance to a deployment claim assumes that synthetic-media classification transfers from controlled benchmark data to real-world use cases. We call this the \textbf{external-validity gap}: whether performance under one evaluation distribution remains informative under deployment conditions~\cite{freiesleben2025benchmarking}. In deepfake detection, this transfer depends on whether the forensic signals that detectors exploit remain reliable across changes in generators, processing pipelines, and transmission channels. Provenance approaches such as content credentials (C2PA~\cite{c2pa2024specification}) attach origin metadata at creation, but this metadata can be stripped by re-encoding or screen capture and establishes origin rather than intent, leaving the consequential-validity gap intact~\cite{openai2024c2pa, worldprivacyforum2024c2pa}.

\begin{figure*}[b]
\centering
\includegraphics[width=0.88\textwidth]{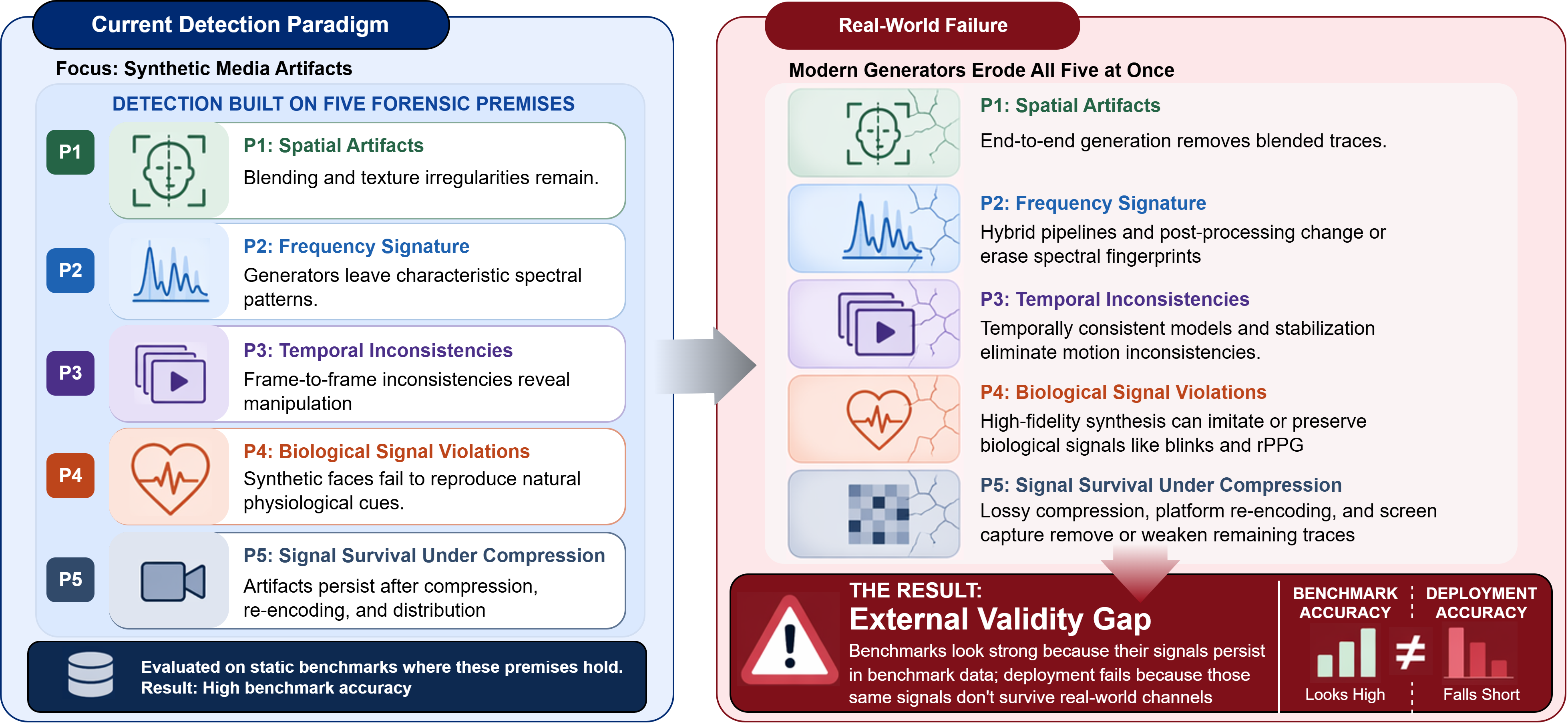}
\caption{Five forensic premises underlying current deepfake detection. Each assumes that synthetic media introduces detectable signals in a specific domain that persist through real-world distribution. Advances in generative models are eroding P1--P4, while real-world distribution undermines P5. These failures produce the external validity gap between benchmark-to-deployment performance. Full detector lists and erosion details appear in Table \ref{tab:premises}.}
\label{fig:forensic-premise}

\end{figure*}

Many deepfake detectors exploit forensic signals that depend on assumptions about the generation process and the preservation of those signals after distribution. We organize these assumptions into five representative forensic premises (Figure~\ref{fig:forensic-premise}). P1–P4 describe spatial, frequency-domain, temporal, and biological cues used by different detector families, while P5 concerns the survival of such cues under real-world processing and transmission. These premises are not universal requirements of all detectors, nor does weakening one imply that a modality is no longer useful. Rather, advances in generation and changes in distribution can undermine particular cues learned from earlier datasets. Consequently, evaluation on curated benchmarks cannot, by itself, establish reliability under unseen generators, pipelines, or communication channels.
\textbf{P2: Frequency signatures} assumes generators leave characteristic spectral fingerprints; hybrid pipelines produce different patterns that older detectors were not trained on
\textbf{P3: Temporal inconsistencies} assumes frame-to-frame flicker and identity drift survive; temporally aware generators and motion stabilization largely remove them
\citep{blattmann2023align,Usmani2025STKD-VViT}.
\textbf{P4: Biological signals} assumes synthetic faces cannot reproduce blink timing, gaze stability, or rPPG patterns; high-resolution generators can now imitate these cues~\citep{Seibold2025Deepfakeheart}.
\textbf{P5: Signal survival under compression} is a meta-premise: it assumes that whatever signal a detector relies on survives lossy compression, social-media re-encoding, screen capture, and conferencing codecs~\cite{Lu2023ImpactOV, chandra2025deepfakeeval}. 
These are not isolated failure modes. A single piece of real-world media distributed over a video-conferencing platform may simultaneously erode P1–P4 through generator advances and P5 through channel degradation. The external-validity gap can therefore compound: benchmark accuracy reflects conditions in which these premises hold, while deployment operates where they do not. A deployment claim requires evaluation across the generator and channel shifts expected in the target setting.

\subsection{Consequential Validity: From Detection to Prevention}
\label{sec:consequential-validity}

\begin{figure*}[t]
\centering
\includegraphics[width=\textwidth]{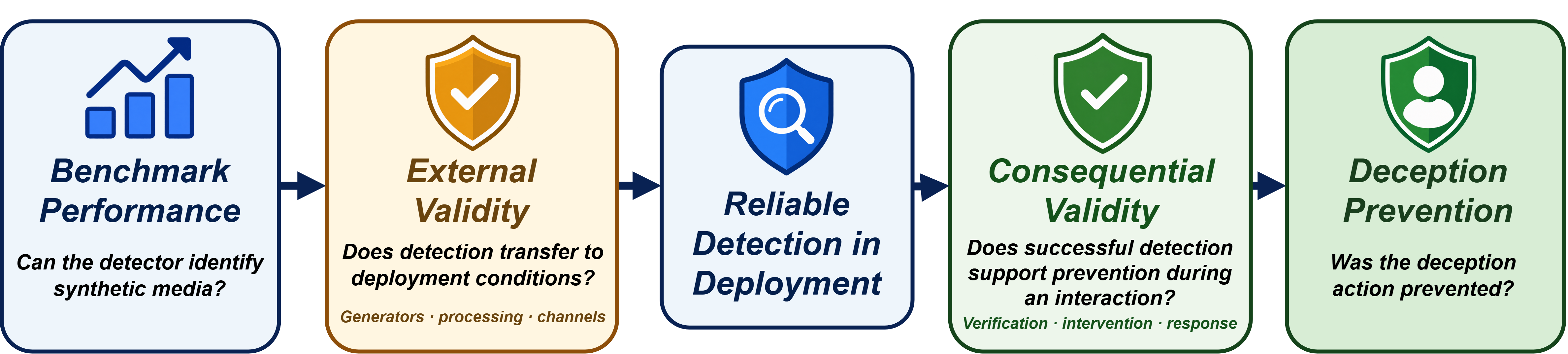}
\caption{Two validity steps between benchmark performance and real-world
deception prevention. External validity asks whether detection transfers to
deployment conditions. Consequential validity asks whether successful detection
supports prevention during an interaction.}
\label{fig:validity-flow}

\end{figure*}

Suppose the external-validity gap (discussed in \S\ref{sec:external-validity}) were fully solved and a detector performed equally well on real-world media. This would show that it can reliably identify synthetic media, but not that deception is prevented. A detector may correctly identify a deepfake yet intervene too late, fail to trigger verification, or fail to change the recipient's decision.
We refer to this second gap as \textbf{consequential validity}, which asks whether successful detection actually helps prevent deception~\cite{freiesleben2025benchmarking}. Prevention depends not only on detection but on when the system intervenes, what action follows, and how the recipient responds. This motivates evaluating the interaction and its outcome, rather than synthetic-media classification alone. We show this flow in Figure \ref{fig:validity-flow}.

\section{A Framework for Evaluating Interactive Deception}
\label{sec:framework}

\begin{figure*}[h]
\centering
\includegraphics[width=\textwidth]{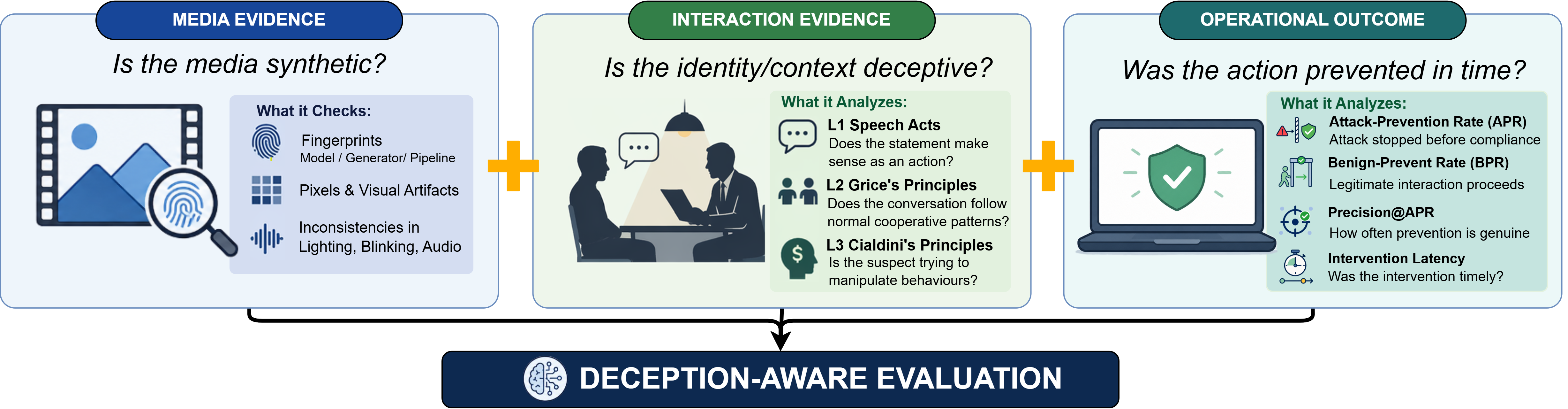}
\caption{Three complementary components of deception-aware evaluation.
Media evidence assesses synthetic origin, interaction evidence assesses
the surrounding communication, and operational outcomes measure whether
the deceptive action is prevented. Their combination provides evidence
for real-world deception prevention.}
\label{fig:deception-component}

\end{figure*}

Section~\ref{sec:validity-gaps} shows that synthetic-origin classification alone does not establish whether interactive deception will be prevented. We define \textbf{interactive deception} as an interaction in which impersonation is used to mislead a recipient about identity, authority, context, or intent in order to influence a decision or action. Evaluating this construct requires evidence beyond synthetic-media classification. We therefore consider three levels of evaluation, summarized in Figure~\ref{fig:deception-component}. The first is \textbf{media evidence (\S\ref{sec:external-validity})}, which asks whether the content is likely synthetic. The second is \textbf{interaction evidence}, which asks whether the identity, context, or behaviour within the interaction provides evidence of deception~\cite{Searle_1979,grice1975logic,cialdini2009influence}. The third is the \textbf{operational outcome}, which asks whether the attack was actually prevented.  A defence may use any subset of these evidence sources to determine whether and when to intervene. We do not prescribe a particular fusion or decision rule; instead, the framework evaluates the resulting chain from evidence, to intervention, to recipient action and outcome.

\subsection{Interaction Evidence}
\label{sec:interaction-evidence}

We propose three interaction-level dimensions for evaluating this construct: \emph{act validity}, \emph{conversational coherence}, and \emph{influence patterns}. Table~\ref{tab:evaluation-components} summarizes these dimensions and their role in the evaluation. They are grounded in Speech Act Theory~\cite{austin1962things,Searle_1979}, Grice's Cooperative Principle~\cite{grice1975logic}, and Cialdini's principles of influence~\cite{cialdini2009influence,cialdini2016pre}. These indicators do not constitute a complete model of deception, but they target observable interaction-level features that media forensics alone cannot capture.

\textbf{L1: Act validity.}
This dimension asks whether the claimed identity, role, or authority is valid~\cite{Searle_1979}. A request to transfer funds, for example, requires confirming that the speaker is authorized to make that request. Independent evidence such as organizational records, callbacks, shared knowledge, or challenge-response verification can support this check. Prior work on speech acts and phishing/BEC communication shows that such signals can be operationalized~\cite{Yu2025SpeechActLLM,ferreira2015analysis}.
\textbf{L2: Conversational coherence.}
This dimension asks whether the interaction is consistent with its prior context and expected communication pattern. Candidate signals include contradictory claims, unusual requests, unexpected topic shifts, or resistance to verification. No single signal is sufficient on its own, since legitimate interactions may also contain unusual requests or contextual changes~\cite{grice1975logic}. Whether these signals reliably distinguish deceptive from legitimate interactions in deepfake settings is an open empirical question that we identify as a priority for follow-up work~\cite{soe2017,lailiyah2025}.
\textbf{L3: Influence patterns.}
This dimension asks whether tactics such as authority, urgency, scarcity, social proof, or reciprocity are used to pressure the recipient toward rapid compliance or discourage verification. No single tactic is inherently deceptive; the diagnostic signal is their co-occurrence and escalation within an interaction, which prior work associates with manipulative intent~\cite{cialdini2009influence}. Initial evidence from social engineering and deepfake persuasion research supports the detectability of these patterns~\cite{zegarow2024,martinek2025}.
 
These three dimensions capture information that matters when appearance and voice can no longer be trusted alone. They complement rather than replace media forensics. Synthetic-origin evidence asks whether the media is likely synthetic, while L1--L3 add evidence about whether the surrounding interaction is deceptive. Operationalizing these dimensions requires observable signals and appropriate reference information: L1 may rely on independent identity or authorization records, while L2 and L3 require annotated interaction-level labels. If human or model judges are used, their reliability should be evaluated.
 
\paragraph{Illustrative annotation.}
To demonstrate that L1--L3 can be grounded in observable features, we provide a worked annotation on a synthetic interaction fragment modeled on reported deepfake fraud patterns~\cite{bloomberg2024ferrari}. Consider a five-turn video-call excerpt in which a caller impersonating a CEO requests an urgent wire transfer from a finance team member.
 
\begin{tcolorbox}[
  colback=gray!8,
  colframe=gray!40,
  boxrule=0.4pt,
  arc=1.5pt,
  left=6pt, right=6pt, top=4pt, bottom=4pt,
  fonttitle=\small\bfseries,
  title=Synthetic interaction fragment: CEO impersonation over video call
]
\scriptsize
\textbf{Turn~1 (Caller):} ``Hi, this is Benedetto Vigna. I need you to process a wire transfer before end of day. It's for a confidential acquisition.''

\medskip
\textbf{Turn~2 (Recipient):} ``Sure, can you send me the approval form through the usual channel?''

\medskip
\textbf{Turn~3 (Caller):} ``We're skipping the normal process on this one. The board wants it done quietly. I'll send the paperwork after.''

\medskip
\textbf{Turn~4 (Recipient):} ``I'd feel more comfortable if I could verify this with the CFO first.''

\medskip
\textbf{Turn~5 (Caller):} ``There's no time for that. If this doesn't go through today, the deal falls apart. I'm counting on you.''
\end{tcolorbox}
 
\noindent\emph{L1 (Act validity).}
Turn~1 asserts CEO identity and claims authority to order the transfer. The caller's request to bypass standard approval (Turn~3) and refusal of independent verification (Turn~5) are negative signals: a legitimate CEO would not ordinarily resist confirmation through established channels.
 
\noindent\emph{L2 (Conversational coherence).}
Turn~3 introduces a deviation from expected organizational process (``skipping the normal process'') and a secrecy condition (``the board wants it done quietly'') inconsistent with standard financial controls. Turn~5 deflects the recipient's verification request rather than addressing it. Both represent violations of Gricean cooperation: the caller's contributions are not responsive to the recipient's reasonable informational needs.
 
\noindent\emph{L3 (Influence patterns).}
The interaction deploys at least three of Cialdini's principles in combination: authority (CEO identity, Turn~1), scarcity and urgency (``no time,'' ``deal falls apart,'' Turn~5), and commitment pressure (``I'm counting on you,'' Turn~5). Their co-occurrence and escalation across turns constitutes the diagnostic pattern; any single tactic alone is common in legitimate business communication.
 
\paragraph{Annotation tractability.}
Each label above is grounded in observable utterance-level features, such as identity claims, process deviations, and temporal pressure language, rather than subjective intent judgments. This suggests that trained annotators or prompted language models could produce these labels with reasonable agreement, though inter-rater reliability on real deepfake interaction data remains to be established empirically.

Table~\ref{tab:codebook} specifies how each indicator could be labelled: the unit of annotation, the observable cues, the reference information required, and a candidate decision rule. These rules are proposals intended to make the indicators testable. Their reliability is an empirical question (\S\ref{sec:limitations}).

\begin{table}[h]
\centering
\scriptsize
\setlength{\tabcolsep}{4pt}
\caption{Proposed annotation codebook for interaction-level indicators.
Candidate decision rules are intended for empirical validation.}
\label{tab:codebook}
\begin{tabular}{@{}p{0.14\linewidth}p{0.14\linewidth}p{0.21\linewidth}p{0.16\linewidth}p{0.25\linewidth}@{}}
\noalign{\hrule height 0.8pt}
\noalign{\vskip 2pt}
\raggedright\textbf{Indicator} &
\raggedright\textbf{Unit} &
\raggedright\textbf{Observable cues} &
\raggedright\textbf{Reference needed} &
\raggedright\textbf{Candidate rule} \tabularnewline
\noalign{\vskip 2pt}
\noalign{\hrule height 0.4pt}
\noalign{\vskip 4pt}
\raggedright\textbf{L1}: Act validity &
\raggedright Scenario: valid / invalid / unverified &
\raggedright Identity or authority claims; requests for transfers, access, or
disclosure &
\raggedright Independent records or out-of-band checks &
\raggedright\emph{Invalid} if independent evidence contradicts the claim;
\emph{unverified} if the claim is not checked or the speaker refuses a check
\tabularnewline
\noalign{\vskip 4pt}
\raggedright\textbf{L2}: Conversational coherence &
\raggedright Turn-level cues; scenario-level count &
\raggedright Contradictions; requests outside normal process; topic shifts;
avoiding verification &
\raggedright Prior history and documented procedures &
\raggedright\emph{Positive} if two or more cue types occur, or if verification
is avoided \tabularnewline
\noalign{\vskip 4pt}
\raggedright\textbf{L3}: Influence patterns &
\raggedright Turn-level tactic tags; scenario-level escalation &
\raggedright Authority; scarcity/urgency; social proof; reciprocity;
commitment; plus secrecy requests &
\raggedright Transcript only &
\raggedright\emph{Positive} if two or more tactics co-occur and intensify,
especially when verification is discouraged \tabularnewline
\noalign{\vskip 4pt}
\noalign{\hrule height 0.8pt}
\end{tabular}
\end{table}

The indicators can overlap. For example, refusing verification affects all three: the claim remains unverified (L1), the conversation becomes less cooperative (L2), and the refusal may be part of a pressure tactic (L3). We therefore treat L1--L3 as complementary evidence, not a scoring system. As discussed in \S\ref{sec:framework}, we do not prescribe how a defence combines them. The scenario-level metrics (\S\ref{sec:evaluation}) measure the outcome of that combination. Indicators can also conflict. A verified identity (L1) combined with genuine time pressure (L3) may indicate a legitimate urgent request. This is why matched benign scenarios are important. Finally, indicators can be assessed before the interaction ends. Cues for L2 and L3 build up as the conversation unfolds, allowing a defence to act before the decision point. Intervention latency and the decision margin $m$, discussed in \S\ref{sec:evaluation}, measure this timing.

\subsection{Evaluating Prevention at the Scenario Level}
\label{sec:evaluation}

If the target construct is interactive deception, clip-level evaluation alone is not enough to support a deployment-level prevention claim. Current benchmarks typically evaluate a media artifact $x$ with ground truth $y \in \{\text{real},\text{synthetic}\}$ and summarize performance using metrics such as accuracy or AUC. For interactive deception, the relevant unit of evaluation becomes the \textbf{interaction scenario} $s$
\[
s =
(\text{channel},
\text{roles},
\text{requested action},
\text{decision point},
\text{termination condition}).
\]
The evaluation question is whether a defence prevents the target action before the relevant decision is made. Moving from a clip to an interaction scenario therefore changes both the unit of evaluation and the definition of success.

\paragraph{Ground truth and outcomes.}
Each scenario should specify whether it is an attack or a benign interaction, the action the attacker is attempting to induce, the decision point, the permitted interventions, and the final outcome. The \emph{decision point} is the point at which the recipient can complete the attacker's requested action. An intervention may take the form of an alert, block, verification request, or escalation.

An intervention counts as prevention only when the harmful action is not completed before the scenario ends. Correctly identifying an attack is therefore not sufficient if the recipient has already complied.
For benign scenarios, the desired outcome is the opposite. Legitimate interactions should proceed without unnecessary intervention. What counts as unnecessary should be defined for the intended deployment setting. For example, an additional verification step may be acceptable in one setting but too disruptive in another.

\paragraph{Decision-aware metrics.}

For a set of attack scenarios $S_{\text{attack}}$ and matched benign scenarios $S_{\text{benign}}$, we propose four measures.

\begin{minipage}[t]{0.45\textwidth}
\vspace{-1em}
\begin{equation}
\text{APR} =
\frac{\bigl|\{s \in S_{\text{attack}} : \text{prevented}(s)\}\bigr|}
{|S_{\text{attack}}|}
\label{eq:apr}
\end{equation}
\end{minipage}
\begin{minipage}[t]{0.52\textwidth}
\vspace{-1em}
\begin{equation}
\text{BPR} =
\frac{\bigl|\{s \in S_{\text{benign}} : \text{pass}(s)\}\bigr|}
{|S_{\text{benign}}|}
\label{eq:bpr}
\end{equation}
\end{minipage}

\textbf{Attack Prevention Rate (APR)} is the fraction of attack scenarios stopped before compliance, as defined in Equation~\ref{eq:apr}. APR requires timely prevention, not correct classification alone. A system that flags an attack only after compliance receives no credit. APR therefore evaluates the broader defence process, including detection, intervention, and recipient response. It should not be reported alone, since a system could achieve high APR by intervening on every interaction.

\textbf{Benign Pass-through Rate (BPR)}, the counterpart to APR, is the fraction of legitimate scenarios allowed to proceed without unnecessary intervention, as defined in Equation~\ref{eq:bpr}. BPR captures the cost imposed on legitimate interactions.

\paragraph{Observed prevention versus defence effectiveness.}
APR measures the proportion of attack scenarios in which
the harmful action is not completed, irrespective of whether
the defence caused that outcome. To estimate the incremental
effectiveness of a defence, evaluations should include a
matched or randomized no-defence control condition.

We propose reporting
\begin{equation}
\Delta \mathrm{APR} =
\mathrm{APR}_{\mathrm{defence}} -
\mathrm{APR}_{\mathrm{control}}
\end{equation}
alongside APR. A causal interpretation requires appropriate
experimental controls. This distinction prevents naturally
unsuccessful attacks from being incorrectly attributed
to the defence.

\textbf{Precision at fixed APR under stated deployment base rates} measures the proportion of interventions that correspond to genuine attacks at an operating point selected to achieve a specified APR. BPR and precision capture different properties. BPR measures how often benign interactions proceed normally, while precision measures how trustworthy an intervention is. Because attacks may be rare relative to legitimate interactions, even a high BPR can still produce many false alerts. Precision should therefore be reported under explicit deployment base-rate assumptions.

\textbf{Intervention latency} measures how quickly the defence acts. The same delay may be early enough in one scenario but too late in another, so latency should be interpreted relative to the decision point. We therefore define the \textbf{decision margin:} $m= t_{\text{decision}} - t_{\text{intervention}}$

where a positive value indicates that the intervention occurred before the decision point.
Together, these measures distinguish clip-level classification ($\text{media}\rightarrow \text{real/synthetic decision} \rightarrow \text{AUC}$) from decision-level evaluation ($\text{scenario}\rightarrow\text{evidence}\rightarrow\text{intervention}\rightarrow\text{recipient action}\rightarrow\text{APR, BPR, precision, latency}$).

AUC remains useful for evaluating the media-classification component. However, it is insufficient as the main evidence for a prevention claim because it averages across operating points, does not account for deployment base rates, and does not show whether the system intervened before the recipient acted.
\begin{table}[h]
\centering
\small
\caption{Hypothetical deployment of 100{,}000 interactions at fixed APR $=$ 90\%. Rows 1, 2, and 4 show the same defence at different base rates; row 3 shows a stronger defence with higher BPR. Precision varies by more than an order of magnitude, while ROC AUC, which is independent of base rate, would remain unchanged for the same underlying detector.}
\label{tab:baserate}
\begin{tabular}{@{}rrrrrr@{}}
\toprule
\textbf{Base rate} & \textbf{BPR} & \textbf{Attacks} & \textbf{Prevented} &
\textbf{Benign interventions} & \textbf{Precision} \\
\midrule
1\%    & 99\%   & 1{,}000 & 900 & 990   & 47.6\% \\
0.1\%  & 99\%   & 100     & 90  & 999   & 8.3\%  \\
0.1\%  & 99.9\% & 100     & 90  & 100   & 47.4\% \\
0.01\% & 99\%   & 10      & 9   & 1{,}000 & 0.9\% \\
\bottomrule
\end{tabular}
\end{table}

\paragraph{Example: Why base rates matter.}
Table~\ref{tab:baserate} shows a hypothetical deployment of 100{,}000 interactions. For simplicity, we assume each prevented attack receives one intervention,
while unprevented attacks receive none. We also assume that each benign interaction that does not pass receives one intervention. At a 0.1\% base rate,
a defence with APR $=$ 90\% and BPR $=$ 99\% prevents 90 attacks but also intervenes on 999 benign interactions. This gives a precision of 8.3\%, or
roughly eleven false alarms for every attack stopped. The same defence has 47.6\% precision at a 1\% base rate and 0.9\% at 0.01\%. ROC AUC would remain
the same across these base rates, even though precision changes substantially. A stronger defence with BPR $=$ 99.9\% reaches 47.4\% precision at the same
0.1\% base rate and APR.

Whether this level of precision is acceptable depends on the cost of an intervention and the harm it prevents. For example, a callback before a high-value transfer may justify more false alarms than blocking routine calls. Two caveats apply. First, BPR measured using deliberately difficult benign scenarios may be lower than BPR on typical traffic. Precision should therefore use BPR estimated from, or adjusted to reflect, the deployment's benign traffic. Second, base rates are rarely known precisely. Precision should therefore be reported across a plausible range, as in Table~\ref{tab:baserate}, and stratified where base rates differ, such as by requested action. Limiting interaction-level checks to high-value decision points can also raise the effective base rate and improve precision.

\paragraph{Scenario-based benchmark.}
A scenario defines the interaction being evaluated. A benchmark should sample scenarios across dimensions needed for coverage, including attack type, recipient role, modality and channel, interaction length, and compliance setup. Each attack scenario should also have matched benign counterparts with similar surface features, such as urgency, authority, or financial requests. This prevents a defence from succeeding simply by relying on obvious cues.

Benchmark size should be determined by the statistical precision required for APR and BPR estimates and by coverage across relevant scenario types, rather than by the number of synthetic clips alone. Reported results should include uncertainty and clearly state the deployment base rate assumed when calculating precision.

APR measures whether prevention was observed, but it does not necessarily measure how much of that prevention was caused by the defence. A recipient may have refused the request even without an intervention. Estimating the added effect of the defence therefore requires an appropriate control or comparison condition, which we treat as a causal-validity requirement.

The goal is not to replace clip-level evaluation but to place it within a broader evaluation protocol: media metrics measure whether synthetic origin can be detected, interaction indicators add evidence of deception, and scenario-level metrics evaluate whether the defence actually prevents attacks without disrupting legitimate ones.


The Tokyo remote-hiring case (\S\ref{sec:introduction}) illustrates this divergence concretely. Media evidence appeared only after the interview; interaction evidence (the applicant's resistance to in-person verification, L2, and a later credential cross-check exposing the false identity, L1) emerged during and after the exchange; and the attack ended before any forensic analysis was conducted. A clip-level benchmark would evaluate whether the recording was correctly classified but would not capture what actually stopped the deception. The full mapping appears in Table~\ref{tab:tokyo-scenario} (Appendix).

\subsection{Building a Scenario-Level Benchmark}
\label{sec:scenario-sourcing}

A scenario-level benchmark requires interaction-length exchanges with ground-truth labels for attack intent, decision points, and outcomes. We identify three complementary sourcing strategies.
(1)  \textbf{Reported incidents.} Public case reports (e.g.,~\cite{unitedpress2026,bloomberg2024ferrari}) provide ecological validity but are sparse, lack full transcripts, and cannot be systematically sampled. They are useful for framework validation but insufficient for statistical evaluation.
(2)
\textbf{Scripted red-team role-play.} Trained operators conduct simulated attacks against participants over real communication channels, controlling scenario dimensions (attack type, channel, recipient role) and producing full transcripts with annotatable decision points. Embedding attack trials within routine interactions improves realism but raises consent requirements. Participant awareness may inflate or deflate prevention rates, so results should report the awareness condition.
(3)
\textbf{LLM-generated scenarios.} Language models can generate diverse interaction transcripts at scale, varying attack type, influence tactics, and recipient behaviour. Generated scenarios can be filtered by human judges for plausibility before inclusion. This approach is scalable but inherits distributional biases from the generating model and may not capture the full range of attacker adaptation. It is best suited for coverage testing and stress-testing edge cases.
A credible benchmark should combine all three: reported incidents for ecological grounding, red-team role-play for controlled measurement, and generated scenarios for coverage. Benchmark documentation should state the sourcing mix, inclusion criteria, and known distributional gaps.

\paragraph{Evaluation protocol.}
We recommend evaluating each scenario under four defence
configurations: (i) no automated defence, (ii) media-forensics-only
detection, (iii) interaction-evidence-only analysis, and
(iv) combined media and interaction evidence.

Each condition should use comparable attack and benign
scenarios with documented recipient roles, evidence availability,
decision points, permitted interventions, and termination
criteria.

Evaluations should record interventions, recipient actions,
and final outcomes separately. APR and BPR should be reported
with uncertainty estimates, while intervention precision
should be calculated under stated deployment base rates.

Comparing the combined defence against the media-only
baseline would establish whether interaction-level indicators
provide incremental operational value.
\subsection{Limitations}
\label{sec:limitations}
Three limitations constrain the applicability
and interpretation of this framework.\textbf{Operational and privacy cost.} 
Interaction-level analysis adds latency and requires access to transcripts, conversational history, or identity records. These costs are uneven. L1 can often rely on existing out-of-band checks (callbacks, challenge questions) without retaining content, whereas L2–L3 require processing conversation content. Deployments must therefore weigh prevention against consent, data minimization, retention, and processing location. Where that access is unavailable, only the media-forensic component of the framework can be applied. \textbf{Cross-cultural transfer.} Conversational norms and influence strategies vary across languages, cultures, and organizational settings. As a result, the proposed Gricean- and Cialdini-based indicators (L2--L3) may raise false positives on benign interactions in unfamiliar settings and should not be assumed to hold outside the settings in which they are validated. BPR should be reported per language and setting, and L1–-L3 should not be assumed to hold beyond where they are validated. \textbf{Empirical grounding.} The framework is conceptual. We do not show that L1–-L3 can be annotated reliably, nor that they improve prevention over a media-forensics-only baseline, and we do not release an instantiated benchmark. The worked annotation in \S\ref{sec:interaction-evidence} shows that labels can be tied to observable features, not that annotators agree on them. The necessary next steps are measuring inter-rater agreement between human and LLM judges on red-team transcripts and comparing APR/BPR against media-only defenses (\S\ref{sec:call_to_action}).

\section{Call to Action}
\label{sec:call_to_action}
Our framework asks for a stronger deployment claim than benchmark accuracy alone, but it does not prove a system will work in practice. That proof has to be built, and it depends on three groups acting together.

\textbf{Researchers and Benchmark Builders}
We ask researchers to change what they evaluate. Benchmarks should move beyond isolated clips to realistic interaction scenarios describing who is involved, what is requested, and when a decision is made, each matched with benign cases sharing the same surface features, such as urgency, authority, and financial requests, so a system must distinguish deception from normal behaviour \cite{raza2026deepfakes}. Researchers should also validate interaction-level indicators rather than assume they work. L1--L3 rely on judgments about identity, coherence, and influence that may vary across people and context. Researchers should test whether these indicators can be labelled reliably, how human and model-based judges compare, and whether the indicators improve on a media forensics-only baseline. Results should be reported carefully. Precision should state the assumed base rate and uncertainty, since rare attacks can produce a high pass-through rate while still hiding many false alerts. 
Finally, attackers will adapt once they learn what a system checks for. Prevention should therefore be tested against new attacks using a fixed defence, while also accounting for realistic recipient behaviour, including people who ignore warnings or act despite their doubts.

\textbf{System and Deployment Designers}
We ask designers to ensure systems can intervene before the recipient acts, designing around the decision margin (m), the time available to intervene, rather than against classification accuracy in the abstract. What helps most at that moment is easy verification rather than a confidence score handed to a user mid-call. The Ferrari attempt was stopped by a shared-knowledge question the impersonator could not answer, and designs that make out-of-band checks simple to invoke, such as callback procedures, challenge questions, and second-channel confirmation for high-value actions, put that kind of check within reach at the point of doubt. And when designers report how their systems perform, we ask them to publish attack prevention and benign pass-through together, at a stated operating point and base rate, since a system that challenges every interaction can post a high prevention rate while making the channel unusable. AUC still says something useful about the classifier, but it averages across operating points and reveals nothing about whether the system acted in time.

\textbf{Policymakers, Procurement, and Standards Bodies}
We ask those who set requirements to match performance claims to the conditions in which they were tested. Claims such as “99\% accurate deepfake detection” should state the generators, channel conditions, operating point, and base rate used in evaluation. Procurement and standards can require this evidence instead of relying on a headline benchmark score. Provenance tools such as content credentials and synthetic-media disclosure can help identify synthetic media, but they do not ensure that deception is prevented. Governance should therefore pair them with verification processes such as callbacks, second-channel confirmation, and dual authorization for high-value actions. Finally, we ask that incident reporting capture what happened before the recipient acted. Reports should record the requested action, when the decision occurred, and whether any intervention took place. This would make it easier to measure prevention across incidents rather than infer it case by case.



\section{Conclusion}

Current deepfake benchmarks measure synthetic-media classification, while deployment claims about interactive deception require evidence that attacks are actually prevented. Two validity links are missing: \textbf{External validity} asks whether benchmark classification performance transfers to deployment media. \textbf{Consequential validity} asks whether synthetic-origin classification provides sufficient evidence for the deployment outcome of deception prevention. Solving the first leaves the second untouched. We therefore propose making the deployment construct explicit, evaluating the interaction scenario rather than the isolated clip, and reporting decision-aware metrics identified in \S\ref{sec:evaluation} under stated conditions. The broader lesson is that a benchmark can be valid for what it measures and still insufficient for the claim built on it.

\bibliographystyle{unsrt}

\bibliography{reference3}

\appendix
\section{Appendix}
\subsection{Core Components of Social-Theoretic Frameworks}
\FloatBarrier
\setcounter{figure}{0}
\setcounter{table}{0}
\renewcommand{\thefigure}{A.\arabic{figure}}
\renewcommand{\thetable}{A.\arabic{table}}
\begin{table*}[hbt!]
\caption{Core components of the three social-theoretic frameworks, their analytical level, and key conditions required for valid communication.}
\label{tab:theory-components}
\centering
\scriptsize
\begin{tabular}{@{}lllll@{}}
\toprule
\textbf{Framework} & \textbf{Level} & \textbf{Component} & \textbf{Description} & \textbf{Key Condition / Signal} \\
\midrule

Speech Act Theory & Individual act & Assertives & Claims about the world & Speaker belief \\
& & Directives & Requests or commands & Authority / standing \\
& & Commissives & Commitments to future actions & Intent to follow through \\
& & Expressives & Expressions of emotion & Sincerity \\
& & Declarations & Change institutional reality & Authority + proper context \\

\midrule

Grice’s Principle & Conversation & Quantity & Appropriate amount of information & Informational balance \\
& & Quality & Truthful contributions & Evidence / truthfulness \\
& & Relation & Relevance to context & Context alignment \\
& & Manner & Clarity and order & Clarity / lack of ambiguity \\

\midrule

Cialdini’s Principles & Recipient response & Reciprocity & Return favors & Obligation \\
& & Commitment & Align with prior commitments & Consistency pressure \\
& & Social Proof & Follow others & Perceived consensus \\
& & Authority & Defer to authority & Status cues \\
& & Liking & Prefer familiar others & Affinity / similarity \\
& & Scarcity & Value limited opportunities & Urgency / time pressure \\
& & Unity & Shared identity & In-group alignment \\

\bottomrule
\end{tabular}
\end{table*}

\FloatBarrier

\subsection{Notation and Terminology Reference}
\label{app:notations_terms}
\begin{table}[h]
\centering
\caption{Glossary of abbreviations used in this paper.}
\label{tab:abbreviations}
\small
\begin{tabular}{p{2.5cm} p{10cm}}
\toprule
\textbf{Abbreviation} & \textbf{Definition} \\
\midrule
APR & Attack Prevention Rate (\S~\ref{sec:evaluation}) \\
AUC & Area Under the Receiver Operating Characteristic Curve \\
BEC & Business Email Compromise \\
BPR & Benign Pass-through Rate (\S~\ref{sec:evaluation}) \\
CNN & Convolutional Neural Network \\
FAccT & ACM Conference on Fairness, Accountability, and Transparency \\
GAN & Generative Adversarial Network \\
LLM & Large Language Model \\
LoRA & Low-Rank Adaptation \\
ML & Machine Learning \\
NeRF & Neural Radiance Field \\
rPPG & Remote Photoplethysmography \\
VLM & Vision-Language Model \\
\bottomrule
\end{tabular}
\end{table}
\FloatBarrier

\newpage
\begin{table}[t]
\centering
\caption{Notation and key terms used in the paper.}
\label{tab:notation}
\small
\begin{tabular}{L{3.0cm} L{9.5cm}}
\toprule
\textbf{Symbol / Term} & \textbf{Definition} \\
\midrule
$s$ & A single interaction scenario being analyzed (e.g., a video call, voice call, or messaging exchange). \\
$S_{\text{attack}}$ & Set of attack scenarios used in benchmark evaluation. \\
$S_{\text{benign}}$ & Set of legitimate interactions that resemble attacks, used to evaluate false-positive resistance. \\
$m$(decision margin) & Time remaining between a system's intervention and the recipient's decision point;$m= t_{\text{decision}} - t_{\text{intervention}}$. An intervention prevents harm only while $m$>0 (\S\ref{sec:evaluation}). \\
Intervention latency & Time elapsed from the start of interaction $s$ to the system's intervention decision (block, alert, or escalation). Correct decisions issued after compliance do not contribute to APR (\S\ref{sec:evaluation}).\\
\addlinespace[0.5em]
\midrule
\addlinespace[0.3em]
External validity & Whether a benchmark's classification performance transfers to deployment conditions such as unseen generators, pipelines, and transmission channels (\S\ref{sec:external-validity}). \\
Consequential validity & Whether successful synthetic-media classification is sufficient for the deployment outcome of deception prevention; the second and deeper of the two validity gaps (\S\ref{sec:consequential-validity}). \\
Precision at fixed APR & The proportion of flagged interactions that are genuine attacks, evaluated at a chosen attack prevention rate and under 
deployment-realistic base rates. Reports how many alerts a defence produces per real attack caught, surfacing the false-positive cost that high BPR alone can hide  (\S\ref{sec:evaluation}).\\
interactive deception & Deception evaluated as a property of an interaction rather than of a media artifact: a communicative act whose validity conditions are deliberately fabricated. \\
Communication analysis & The complementary analytical layer proposed in this paper, which targets behavioural and pragmatic signals beyond the reach of artifact-based forensics.\\
Validity conditions & The contextual, situational, and intentional requirements that must hold for a speech act to be sincerely and successfully performed (Speech Act Theory). \\
Layer 1 (L1) & Illocutionary: act-validity analysis at the utterance level (Speech Act Theory). \\
Layer 2 (L2) & Conversational norm analysis at the conversation level (Grice's cooperative principle). \\
Layer 3 (L3) & Perlocutionary: coercion-pattern analysis at the recipient-response level (Cialdini's principles of influence). \\
Artifact & A detectable trace left by a generative process in synthetic media (e.g., blending boundaries, frequency anomalies, temporal flicker); the primary target of current detection methods. \\
Media forensics & The analysis of audiovisual signals for evidence of synthetic origin, encompassing artifact-based detection methods. \\
Base rate & The proportion of attack interactions among all interactions in a given deployment context. Deepfake fraud has very low base rates 
(typically far below 1\%), which makes false-positive volumes operationally significant even when BPR is high (\S\ref{sec:evaluation}).\\
Operating point & A specific decision threshold at which a classifier or defence is configured to operate, determining the trade-off between 
true-positive and false-positive rates. Deployment systems operate at one 
chosen point rather than averaging over all possible thresholds.\\
Phishing & Deceptive electronic communication (typically email) designed to obtain sensitive information or induce harmful actions by impersonating a trustworthy entity.\\
Red-team & An adversarial evaluation protocol in which a separate team of attackers attempts to defeat a frozen defence system, providing an ecological validity check against attacker adaptation.\\
Vishing & Voice phishing: social engineering conducted over telephone or voice channel, including attacks that use voice cloning.\\
\bottomrule
\end{tabular}
\end{table}

\newpage
\FloatBarrier
\begin{table}[h]
\centering
\caption{The five forensic premises underlying current deepfake detection (Section~\ref{sec:external-validity}). Each row gives the assumed signal, the detection methods built on that assumption, and the technical or deployment shifts that erode it. P5 is a meta-premise that conditions the observability of P1--P4 at deployment.}
\label{tab:premises}
\scriptsize
\begin{tabular}{p{0.6cm} p{0.6cm} p{1.7cm} p{2.8cm} p{2.6cm} p{2.6cm}}
\toprule
\textbf{Premise} & \textbf{Period} & \textbf{Signal Type} & \textbf{Underlying Assumption} & \textbf{Representative Detectors} & \textbf{Why It Is Eroding} \\
\midrule
P1 & 2017--2022 & Spatial artifacts & Synthetic faces leave visible traces at blending boundaries and in local textures. & XceptionNet~\cite{rossler2019faceforensics}, Face X-ray~\cite{li2020face}, LAA-Net~\cite{nguyen2024laa} & End-to-end diffusion synthesizes entire frames; no blending step exists~\cite{rombach2022high, yan2024df40}. \\
\addlinespace
P2 & 2019--2022 & Frequency signatures & Generators leave recognizable spectral fingerprints (e.g., checkerboard patterns, anomalous distributions). & F3-Net~\cite{1Qian2020ThinkingFF}, FreqNet~\cite{tan2024frequency}, FE-CLIP~\cite{gong2025feclip} & Hybrid pipelines and post-processing alter or erase prior fingerprints~\cite{corvi2023detection, ricker2024aeroblade}. \\
\addlinespace
P3 & 2019--2023 & Temporal coherence & Sequential generation produces frame-to-frame flicker, identity drift, and unnatural motion. & FTCN~\cite{Zheng2021FTCN}, AltFreezing~\cite{wang2023altfreezing}, MSVT~\cite{Wang2023MVST}, TCN~\cite{Amin2024TempCohereDF} & Temporally aware generators, motion stabilization, and interpolation eliminate these artifacts~\cite{blattmann2023align,Usmani2025STKD-VViT}. \\
\addlinespace
P4 & 2018--2023 & Biological signals & Synthetic faces fail to reproduce blink timing, gaze stability, and rPPG patterns. & Blink-based methods~\cite{li2018InIctuOculi, jung2020deepvision}, gaze models~\cite{Demir2021gaze}, FakeCatcher~\cite{ciftci2020fakecatcher}, DeepRhythm~\cite{Hua2020deeprhythm} & High-resolution generators can preserve or imitate physiological cues~\cite{Seibold2025Deepfakeheart}. \\
\addlinespace
P5 & 2018--present & Signal survival (meta-premise) & P1--P4 signals survive compression, re-encoding, screen capture, and conferencing codecs. & \textit{Conditions all P1--P4 detectors at deployment.} & Detectors trained on clean data; sharp degradation under realistic transmission conditions~\cite{hussain2021adversarial}. \\
\bottomrule
\end{tabular}
\end{table}

\begin{table}[h]
\centering
\caption{Evaluation components for interactive deception.}
\label{tab:evaluation-components}
\small
\begin{tabular}{lll}
\toprule
\textbf{Component} & \textbf{What is measured} & \textbf{Evaluation question} \\
\midrule
Media forensics
& Synthetic-origin evidence
& Is the media likely synthetic? \\

L1: Act validity
& Identity, authority, role
& Is the claimed role valid? \\

L2: Conversational coherence
& Context and discourse
& Does the interaction fit its context? \\

L3: Influence patterns
& Compliance pressure
& Is the recipient being pressured? \\

Operational outcome
& Prevention and latency
& Was deception prevented? \\
\bottomrule
\end{tabular}
\end{table}

\FloatBarrier
\subsection{Illustrative Example: Detection vs. Prevention in Remote Hiring (Tokyo, 2026)}
\label{subsec:case-tokyo}

\begin{table}[h]
\centering
\caption{Scenario decomposition for the Tokyo remote-hiring case. Entries
distinguish facts on the public record from class-level mappings; the outcome
is left indeterminate where the reporting does not resolve it~\cite{unitedpress2026}}
\label{tab:tokyo-scenario}
\scriptsize
\renewcommand{\arraystretch}{1}
\begin{tabular}{@{}p{0.16\linewidth} p{0.36\linewidth} p{0.42\linewidth}@{}}
\toprule
\textbf{Layer} & \textbf{What is observed} & \textbf{Mapping / status} \\
Scenario ($s$) &
\emph{channel}:Remote video interview; \emph{roles}:attacker impersonates a real IT executive; recipient is the hiring team &
\emph{requested action}: extend an offer; \emph{Decision point}: the hire / onboarding decision \\

Media forensics &
Hairline boundary, eye misalignment, and lip--audio mismatch were assessed post-hoc, off recorded footage, by third parties &
Consistent with P1 (boundary) and P3 (temporal) classes \S\ref{sec:external-validity} argues are eroding; detector \emph{out of the decision loop} \\

Interaction (L1--L3) &
Applicant requested fully remote work and ended the interview when told that in-person attendance was required; the resume was later found to match a real executive &
Later credential cross-check exposed the false identity (L1). Withdrawal when an in-person verification condition was introduced provides evidence of resistance to verification (L2). L3: not assessable from the public record.\\

Outcome &
Attacker withdrew before any forensic analysis &
Decision margin $m$ not reported (\S\ref{subsec:case-tokyo}) \\
\bottomrule
\end{tabular}
\end{table}

The divergence between synthetic-media detection and prevention appears in a reported remote-hiring case (\S\ref{sec:introduction})~\cite{unitedpress2026}. An applicant used AI to impersonate a real executive during an online interview at a Japanese IT company, submitting credentials later found to belong to that person. Outside analysts assessed the footage as likely AI-generated only after the interaction. We use this case to illustrate how synthetic-media detection and deception prevention can diverge, and how our framework captures information that clip-level evaluation does not. Table \ref{tab:tokyo-scenario} maps the case to our framework.

The case contains three distinct forms of evidence and outcome. \textbf{Media evidence} appeared after the interview, when analysts identified signs that the recording was likely synthetic. \textbf{Interaction evidence} appeared during and after the exchange: the applicant requested fully remote work and withdrew when in-person attendance was required, providing an L2-relevant response to verification, while a later resume check provided L1 evidence that the claimed identity was false. \textbf{Operationally}, the interview ended without a hire before the forensic assessment was conducted.

This decomposition illustrates the consequential-validity gap. A clip-level benchmark could evaluate whether the recording was correctly identified as synthetic, but it would not show whether that evidence helped prevent the deception, what other evidence affected the interaction, or what happened before the outcome. Scenario-level evaluation makes these distinctions explicit by evaluating the media evidence, interaction evidence, and final outcome separately.

The case does not show how often these mechanisms occur. It also provides no interview transcript, so L2 can only be mapped from the reported sequence of events and L3 cannot be assessed. Even with these limits, the case illustrates why synthetic-media detection and deception prevention should be evaluated separately.

\end{document}